\documentclass[linenumbers, twocolumn]{aastex631}

\shorttitle{modeling AGBs in Monash and MESA}
\shortauthors{Cinquegrana, Joyce and Karakas}
\usepackage{chngcntr}
\usepackage{ulem}

\begin{document}
\nolinenumbers

\title{Bridging the Gap between Intermediate and Massive Stars I: Validation of MESA against the State-of-the-Art Monash Stellar Evolution Program for a 2$M_{\odot}$ AGB Star}

\author[0000-0001-7902-8134]{Giulia C. Cinquegrana}
\affiliation{School of Physics \& Astronomy, Monash University, Clayton VIC 3800, Australia\\
}
\affiliation{ARC Centre of Excellence for All Sky Astrophysics in 3 Dimensions (ASTRO 3D) \\
}

\author[0000-0002-8717-127X]{Meridith Joyce}
\affiliation{Space Telescope Science Institute, 3700 San Martin Drive, Baltimore, MD 21218, USA\\
}
\affiliation{ARC Centre of Excellence for All Sky Astrophysics in 3 Dimensions (ASTRO 3D) \\
}

\author[0000-0002-3625-6951]{Amanda I. Karakas}
\affiliation{School of Physics \& Astronomy, Monash University, Clayton VIC 3800, Australia\\
}
\affiliation{ARC Centre of Excellence for All Sky Astrophysics in 3 Dimensions (ASTRO 3D) \\
}

\begin{abstract}
\nolinenumbers

One--dimensional stellar structure and evolution programs are built using different physical prescriptions and algorithms, which means there can be variations between models' predictions even when using identical input physics. This leads to questions about whether such deviations are physical or numerical; code validation studies are important and necessary tools for studying these questions. We provide the first direct comparison between the Monash stellar evolution program and MESA for a $2M_{\odot}$ model evolved from the zero-age main sequence to the tip of the thermally pulsing asymptotic giant branch. We compare the internal structure of the two models at six critical evolutionary points and find that they are in excellent agreement in characteristics like central temperature, central density and the temperature at the base of the convective envelope during the thermally pulsing asymptotic giant branch. The hydrogen-exhausted core mass between the models differs by less than 4.2\% throughout the entire evolution, the final values vary only by 1.5\%. Surface quantities such as luminosity and radius vary by less than 0.2\% prior to the asymptotic giant branch. During thermal pulses, the difference extends to 3.4\%, largely due to uncertainties in mixing and the treatment of atmospheric boundary conditions. Given that the veteran Monash code is closed source, the present work provides the first fully open-source computational analog. This increases accessibility to precision modeling on the asymptotic giant branch and lays the groundwork for higher-mass calculations that are performed with MESA but preserve the standards of the Monash code during the AGB.
\end{abstract}



\section{Introduction} \label{sec:intro}

Due to the practical limitations of computing, stellar evolution codes are typically built to specialize in one mass range or a particular phase of evolution. For example, stellar evolution programs that specialize in the low- to intermediate-mass regime (between $\approx 1$--$8M_{\odot}$) include ATON (\citealt{Ventura98, Ventura13}), Monash (adaptation of the Mount Stromlo Stellar Evolution code; \citealt{LattanzioThesis, Lattanzio86, Frost96, Karakas07}), GARSTEC (the Garching Stellar Evolution Code; \citealt{Weiss08garstec}), EVOL (\citealt{Blocker95a, Herwig04dredge, Herwig04nuclear}), DSEP (the Dartmouth Stellar Evolution Program; \citealt{Dotter07acs}), Cambridge STARS \citep{Eggleton71evolution}, and YREC (the Yale Rotating Stellar Evolution Code; \citealt{Demarque08yrec}). Some of these, including STARS and Monash, evolve through C burning (and further, in the case of Ne burning functionality introduced in some versions of the STARS code), but most are optimized for low-mass calculations. The BaSTI program (a Bag of Stellar Tracks and Isochrones; \citealt{Pietrinferni04large, Hidalgo18updated, Pietrinferni21updated, Salaris22updated}) and PARSEC (the PAdova and TRieste Stellar Evolution Code; \citealt{Bressan12}) extend from the low-mass range to $15 M_{\odot}$ and $12 M_{\odot}$, respectively, but halt evolution at the onset of carbon ignition. While a version of the Monash code has been used to produce super AGB models (7.7 to 10.5$M_{\odot}$) that were validated against the STAREVOL models \citep{Siess07, Doherty10super} in the past, we do not have access to this version presently.

Programs that model evolution past carbon burning include FRANEC (the Frascati Raphson Newton Evolutionary Code; \citealt{Chieffi98evolution, Limongi06nucleosynthesis}), KEPLER (\citealt{Weaver78presupernova, Fuller86evolution, Woosley07nucleosynthesis, Woods20monolithic}), TYCHO \citep{Young01observational} and the Geneva stellar evolution code (\citealt{Eggenberger08geneva}). 

GENEC has been used previously to calculate the evolution of low-mass stars \citep[e.g.][]{Charbonnel96grids}, but these sequences only extend to the end of the early asymptotic giant branch (AGB). FRANEC differs from the aforementioned programs in that it is also capable of modeling the thermally pulsing AGB phase \citep[e.g.][]{Cristallo07molecular}. The stellar programs mentioned above are some of the most commonly cited in the literature, but this is by no means an exhaustive list of all current software in use. 

The limitations of the philosophy of specialization are most apparent when interesting stellar physics happens on the borders of these mass ranges. In \citet{Karakas22paper1} and \citet{Cinquegrana22paper2}, we used the Monash stellar evolution code and post-processing nucleosynthesis code to model the evolution and chemical contributions of super metal-rich, low- and intermediate-mass AGB stars. The Monash code was originally built to model low-mass stars and has been adapted over some years to model intermediate-mass AGBs \citep{Faulkner68evolution, Faulkner72thermal, Gingold74asymptotic, Wood81helium, Lattanzio89carbon, Frost96, Lattanzio92hot, Frost97brightest, campbellThesis}. However, with increasing initial mass (especially at super-solar metallicities) the code struggles to converge through the entire thermally pulsing phase and can cut the evolution of the model prematurely. In particular, peculiarities emerged in 6.5 to 8$M_{\odot}$ models at $Z=0.09, 0.10$ (see Figure 6 in \citealt{Karakas22paper1}). In \citet{Karakas22paper1}, we found that carbon ignition was occurring at a lower initial mass than expected, for the highest metallicity models. To confirm that this is a physical, rather than numerical, phenomenon
requires either significant alteration of the Monash code or attempting to replicate these results with another code, such as MESA, which has extended capability in this mass range. 

In comparison to Monash, the Modules for Experiments in Stellar Astrophysics (MESA; \citealt{Paxton10instrument1, Paxton13instrument2, Paxton15instrument3, Paxton18instrument4, Paxton19instrument5}) stellar evolution program has a shorter usage history, but it is unique for a few reasons. First, it is a fully open-source tool, available to any researcher that has the computing capability. Second, MESA has been developed to model almost the entirety of the stellar evolution process for a wide range of masses. This capability is nearly unique in the modeling landscape, making MESA the obvious candidate to bridge the gaps in initial mass not covered by codes that model only low- and intermediate-mass stars well, like Monash, or only high-mass stars, like KEPLER. 

An example of a situation in which MESA was used effectively to supplement specialized calculations from other codes is the work of the NuGrid group, who aim to provide a self-consistent stellar yield database for a wide range of initial masses and metallicities.
In their first release, \citet{Pignatari16} use a combination of MESA and GENEC to produce their models. Given that GENEC is not designed to simulate thermally pulsing AGBs, MESA provided an important supplement. Evolutionary tracks from both were then fed into their multi-zone post-processing nucleosynthesis code (MPPNP; \citealt{Herwig08nucleosynthesis, Pignatari12nugrid}), to produce stellar yields. In their later releases \citep{Ritter18, Battino19, Battino22nugrid}, MESA is used to evolve all updated models. 

Substantial advances in stellar astrophysics are not the result of one researcher using one particular stellar program, but the iteration of many that find the same peculiarities. The outputs of evolution programs are not solely dependent on the input physics we specify, so it is important to compare different software instruments where possible to ensure that we are correctly identifying features we observe in their output as science, rather than mis-classifying code-specific numerical artifacts as physical. These comparisons have a long standing history in the field of astrophysics. For example, \citet{Bahcall92standard}, who validated a strong agreement between solar neutrino calculations of various programs at that time.

Since MESA's relatively recent inception, there have been a variety of validations comparing MESA against the results of EVOL, GENEC, DSEP, BaSTI, FRANEC, GARSTEC and KEPLER. For example, \citet{Paxton10instrument1} provides comparisons of MESA for low--mass models ($0.8$--$2 M_{\odot}$) with numerous tracks provided by \citet{weiss_cassisi_dotter_han_lebreton_2006} and \citet{Herwig04nuclear}. They also validate MESA at higher masses with a $25 M_{\odot}$ comparison against KEPLER and FRANEC. \citet{Jones15codecomp} compares massive models ($15, 20, 25 M_{\odot}$) between MESA, GENEC and KEPLER and identifies the downstream impact the various programs have on nucleosynthesis (using MPPNP). \citet{Sukhbold14compactness} compares an even higher mass range (15 to 65$M_{\odot}$), specifically identifying the numerical impact on the compactness of a pre-supernova core. \citet{Martins13comparison} calculate massive stellar models with MESA and STAREVOL \citep{Forestini94low, Siess97synthetic, Siess00internet} and compare these tracks against both GENEC, STERN \citep{Brott11rotating}, PARSEC and FRANEC as well as observations of Galactic stars. Recently, \citet{Aguirre20aarhus} compares red giant branch models between nine different evolution codes, including MESA. Further comparisons with MESA are published in \citet{Agrawal22explaining, Campilho22atomic, Van123he}. 

In this work, we provide the first validation in the literature of MESA (version 15140; \citealt{Paxton19instrument5}) against the Monash code. In particular, we focus on the thermally pulsing AGB. This is a notoriously difficult phase of stellar evolution to model because the TP-AGB is highly sensitive to both choice of input physics and numerical variations between different programs. There has been little in the way of formal comparison published between software with this capability for a few reasons. Firstly, MESA is the only actively maintained, fully documented, and fully open-source stellar evolution program available; the programs mentioned in the preceding paragraphs are all closed-source\footnote{They are all ``closed-sourced'' in the sense that they do not operate under an open source license. While in some cases the code itself is available openly or by request, the level (or lack) of documentation makes use by a non-expert extremely difficult.} instruments. Secondly, each of these software programs varies widely in their choice of input physics. This increases the difficulty of classifying deviations between results, because it is not immediately clear which variations are caused by input physics as opposed to differences in the algorithms and/or frameworks of the programs themselves. 

One of the most impactful variations in input physics is the treatment of convection. For example, the full spectrum of turbulence model \citep{Canuto91, Canuto96} is used in ATON \citep{Ventura98, Ventura20} as opposed to the Mixing Length Theory of Convection (MLT; \citealt{Prandtl25, Vitense53}) used in the Monash and MESA software. Within MLT, some programs use a constant, solar-calibrated value of~$\alpha_{\rm MLT}$ \citep[such as in this work; see][]{Cinquegrana22solarcal}. Others use $\alpha_{\rm MLT}$ values adapted to evolutionary phase and calibrated to observations more metal-poor than the Sun \citep[see][]{Joyce18not, Joyce18class} or adaptive values of $\alpha_{\rm MLT}$ that are entropy calibrated, used in modified versions of YREC \citep{Spada19testing}. 

Methods to determine the boundaries of convective regions include the predictive mixing and convective pre-mixing algorithms available in MESA \citep{Paxton18instrument4, Constantino15tre, Bossini15unc}. We find these do not result in third dredge up without forcing exponential overshoot (or convective boundary mixing; \citealt{Herwig00}). In Monash, we use the \textit{relaxation} method defined in \citet{Lattanzio86}, which results in significant third dredge up with no forced overshoot down to a minimum mass dependent on initial metallicity (1.5$M_{\odot}$ at $Z=0.004$, 3.5$M_{\odot}$ at $Z=0.04$; \citealt{Karakas10Up, Karakas22paper1}). However, modified versions of the Monash code do include the option of step overshoot to increase the efficiency of the third dredge up, given that observations suggest this process should occur in stars with lower initial masses than we find in our models \citep[see][]{Karakas10mixing, Kamath12}. In STAREVOL, they rely on the Schwarzschild convective stability criterion, which extends no extra mixing beyond the convective boundary and results in no third dredge up \citep{Siess10evolution}. 

Another important variation is the treatment of mass loss and stellar winds. Most programs utilise the \citet{Reimers75} mass loss approximation for the first giant branch; however, it is debated whether the approximations of \citet{Vassiliadis93} or \citet{Blocker95a} are more appropriate for the AGB. Even when using the same AGB mass loss approximation, authors use wildly different scaling factors. For example, an $\eta _{\text{Bl\"{o}cker}}$ value of unity was recommended in the original \citet{Blocker95a} paper. \citet{Kalirai14core}, however, uses values of $\eta _{\text{Bl\"{o}cker}} \approx 0.2$ in their models. \citet{Ventura12dust} uses values closer to $\eta _{\text{Bl\"{o}cker}}=0.02$, which were calibrated against low-metallicity populations of the large Magellanic Cloud \citep{Ventura99, Ventura01lithium}; we do similarly in Monash. Other research teams vary this parameter as a function of evolutionary phase. \citet{Pignatari16} uses values of $\eta _{\text{Bl\"{o}cker}}=0.01$ for the oxygen-rich phase, but increases the parameter up to $\eta _{\text{Bl\"{o}cker}}=0.08$ depending on mass, for models that reach C / O $> 1$. 

The wide variance in the mass loss scaling factor alone, despite using the same mathematical formalism, indicates that there are also significant differences in how each program defines their atmospheric boundary conditions, which directly dictate the impact of the mass-loss efficiency. We also use different nuclear networks and reaction rates (though the bulk of reaction rates are generally sourced from JINA REACLIB; \citealt{Cyburt10} or STARLIB; \citealt{Sallaska13starlib}). The opacity tables of \citet{Iglesias96} are frequently used for high temperature regions, but low temperature opacities vary between purely solar and solar-scaled molecular opacities \citep[e.g.][]{Alexander94low, Ferguson05low} to the more recently built CNO and $\alpha$-enhanced molecular opacities of \citet{Lederer09low} and the \AE SOPUS software tool of \citet{Marigo09}. Further complicating matters, programs vary in their use of analytical integration schemes for atmospheres \citep[e.g.][]{Eddington26, Krishna66} versus the attachment and interpolation of tables of model atmosphere calculations from 3D hydrodyanmical simulations, like those of PHOENIX \citep{Hauschildt99nextgena, Hauschildt99nextgenb}. 

Hence, the huge extent to which stellar evolution programs with AGB modeling capabilities vary in their frameworks and physics makes meaningful and quantitative comparisons of their results difficult. It is important, however, to mention that the major reason for inconsistencies in the choices made for input physics discussed above is our incomplete understanding of how complex AGB processes, like the third dredge up, physically work. This in itself demonstrates the importance of comparing programs. To achieve a better understanding of how our frameworks differ and the propagating impacts code differences have on scientific predictions, we aim to address this gap in the knowledge for two of the major codes used within this field. 

We compare the evolution of a 2$M_{\odot}$, $Z=0.03$ model computed with MESA to the same computed with the Monash code (and include an identical 7$M_{\odot}$ comparison in the appendix). Our methods and input physics are detailed in \S~\ref{sec:methods}. In \S~\ref{sec:results}, we validate the consistency of the internal structure between the models at six critical points in evolution and provide a general comparison of the surface properties. Finally, discussion of our results and concluding remarks are given in \S~\ref{section:discussionandconclusion}.

\section{Methods} \label{sec:methods}

Rather than trying to clone the algorithms and numerics of the Monash code in MESA verbatim, the purpose of this work is to provide the settings and control physics that produce an AGB model that is structurally consistent between both codes. In the following subsections, we begin with a discussion of the differences in the structure and organization of the source codes. We then describe each component of the input physics (i.e., adjustments than can be made with user controls) and the accuracy of our match.  

\subsection{Numerics}

The Monash stellar evolution program and MESA differ in their organization, structure, and numerics in a few ways. Both are written in Fortran and model the temporal behavior and structure of a polytrope. In both programs, full structural output is available at any time step, with the output frequency set by the user, and a global evolutionary record is updated once per evolutionary step. These outputs include a combination of human-readable data files and Fortran binary files. While libraries of user controls pertaining to output customization (\texttt{profile\_columns} and \texttt{history\_columns} lists) are available in MESA, there is no such analog in Monash---a user would need to update the source code directly to modify the quantities recorded in output. Both programs adopt a variant of the Newton--Raphson iterative scheme to solve the matrix decomposition representing the stellar structure equations (in Monash, the Henyey matrix numerical method as described in \citealt{Henyey59} and \citealt{Henyey64}). In MESA, global surface quantities, such as $T_{\text{eff}}, R_{\text{star}}, \log L_{\text{star}}$, represent surface averages over the whole model taken at the outer-most structural cell. The convention in Monash is the same. 

Both programs are thread-safe\footnote{Multiple instances can be run on the same computer simultaneously without data corruption, so long as the output location is different for each instance.} and require a Fortran compiler; while this is the only external system requirement for Monash, all components of the MESA Software Development Kit (SDK; \citealt{richardtownsend2019dash2669541}) are required to run MESA. The Monash code is not parallelized. While some elements of MESA are parallelized, its performance does not improve above roughly 18 threads. The number of threads assigned to a MESA instance can be set by the environment variable \texttt{OMP\_NUM\_THREADS}. While both programs can be run on a standard 2- or 4-core laptop, we have found that devoting at least 8 threads to MESA calculations improves performance.

In Monash's case, the creation and adjustment of a ZAMS model is a procedure handled separately from the evolution of that model, whereas this functionality is included by default in a standard MESA evolutionary run and can include directly computing the pre-main sequence evolution where desired. There are of order 10--20 user control parameters in Monash that are routinely adjusted, with the option to change up to of order 100--200, whereas specialized MESA \texttt{inlists} can include dozens to hundreds (selected from a library of tens of thousands). 

As Monash is specialized for low- and intermediate-mass stars, it contains a much smaller code base and runs faster than MESA by more than a factor of 10 (see Table \ref{table:2} for a comparison of run times for our $2M_{\odot}$ case; in the $7M_{\odot}$ case, the run time difference favors Monash by more than a factor of 20). Monash is also quite fast relative to other stellar evolution programs that run on single cores. Not unrelated, the Monash code is also exceptionally light in its memory use (as it was written in an era where such resources were not abundant), whereas MESA uses a large amount of memory.
Overall, the Monash stellar evolution code is compact, closed-source, fast, low-memory, highly specialized, and has a small, expert user base over which different versions are distributed (that is to say, no ``main'' version or branch is centrally maintained). In contrast, MESA is large, flexible, open--source, slower, high-cost in memory, maintained on a centralized, open-access platform (GitHub), has exceptionally broad coverage of the stellar mass and metallicity regimes, and has a large and active user base comprising experts with a diverse array of astrophysical specializations as well as novices.

Key milestones in the recent development of the Monash code are described in detail in \citet{Karakas10} and \citet{Kamath12}---in particular, the introduction of convective overshoot to accompany the instantaneous mixing approximation for convection. For a more complete discussion of the algorithms used in Monash not covered in the following sections, we refer the reader to \citet{LattanzioThesis, Lattanzio86, Frost96, Karakas07, campbellThesis}. However, we note that the version of the code used in this work assumes instantaneous mixing in convective zones \footnote{The assumption of instantaneous mixing in convective regions is sufficient for this mass range and acts as a decent approximation to hydrodynamical models of convection. In particular, it is a reasonable approximation to the turbulent convection experienced by models during helium shell flashes \citep{Stancliffe2011three} in terms of convective velocities and turnover times. More problematic is the treatment of convective borders. This is discussed further in Section. \ref{convection}.}, as opposed to the version described in \citet{campbellThesis, Campbell2008evolution},\footnote{\citet{campbellThesis} is available for download at \url{https://users.monash.edu.au/~scamp/downloads/phd-thesis-Campbell.pdf}} which treats convection as diffusive. For a thorough discussion of MESA's numerics, see \citet{Paxton10instrument1}, and see subsequent MESA instrument papers for a historical record of extensions to MESA's functionality.


\subsection{Initial quantities}

In both codes, we calculate the initial helium abundances based on the initial metallicity of the model: 
\begin{equation}
    \rm{Y}_i = Y_0 + \frac{\Delta Y}{\Delta Z} \times \rm{Z}_i.
\end{equation}
The primordial helium abundance, $Y_0$, provides a lower limit on the initial helium abundance in the first stars \citep{Karakas14}. It has an observed value of $Y_0=0.2485$ \citep{Aver13}. Over time, the helium abundance increases at the rate of the helium-to-metal enrichment ratio, $\frac{\Delta Y}{\Delta Z}=2.1$ \citep{Casagrande07}. So, with an initial metallicity of $\rm{Z}_i=0.03$, we calculate $\rm{Y}_i=0.30$. 

\subsection{Equation of state}

The equation of state (EOS) describes the relationship between the pressure, density and temperature of a plasma. The EOS in the Monash code is based on the formulae of \citet{Beaudet67}, \citet{Baerentzen65}, the Saha equation and the ideal gas law with radiation pressure term. The MESA EOS utilizes the work of OPAL \citep{Rogers2002}, SCVH
\citep{Saumon1995}, FreeEOS \citep{Irwin2004}, HELM \citep{Timmes2000},
and PC \citep{Potekhin2010}.

\subsection{Nuclear network and reaction rates}

The nuclear network in Monash explicitly follows six isotopes: $^{1}$H, $^{3}$He, $^{4}$He, $^{12}$C, $^{14}$N and $^{16}$O. These isotopes are contained within the \verb|basic.net| nuclear reaction network option in MESA. Given that this work is an accompanying result of Cinquegrana, Joyce and Karakas (2022, \textit{in prep}) in which we model massive stars, our science case requires that we adopt a more extended network in MESA. To accomplish this, we use the function \verb|auto_extend_net| to switch between three nuclear networks based on which evolutionary process the model is going through. These are \verb|basic.net|, \verb|co_burn.net| and \verb|approx21.net|. Together, these networks cover hydrogen burning through silicon burning. The reaction rates in MESA are sourced from JINA REACLIB \citep{Cyburt10}, with some additional weak rates from \citet{Fuller85stellar, Oda94rate, Langanke00shell}. Monash predominantly adopts the rates of \citet{Harris83thermonuclear} for hydrogen burning (or \citet{Fowler75thermonuclear} for rates that were not updated in \citet{Harris83thermonuclear}). For helium and carbon burning, Monash uses rates from \citet{Caughlan88}.

\subsection{Convection}
\label{convection}

The default treatment of convection in both Monash and MESA is the Mixing Length Theory of convection (MLT; \citealt{Prandtl25, Paxton10instrument1}). There are multiple prescriptions of the MLT available in MESA; we use \citet{Henyey65mlt}. The \citet{Henyey65mlt} formulation is a slightly modified version of the work of \citet{Vitense53}, who first applied the MLT formalism to stellar modeling. This prescription includes super-adiabatic regions that occur in stellar envelopes. Given the sensitivity of $\alpha_{\rm MLT}$ to both numerics and input physics, we performed an explicit solar calibration of $\alpha_{\rm MLT}$ to determine an appropriate value for our science in MESA. In \citet{Cinquegrana22solarcal}, we determined $\alpha_{\rm MLT}$ by reproducing the solar luminosity and radius to a precision such that further refinement no longer reduced the residuals in luminosity and radius. We found $\alpha_{\rm MLT}=1.931 H_P$ was the most suitable for our physics. Solar calibrations have previously been performed for the Monash code, where a value of $\alpha_{\rm MLT}=1.86 H_P$ was deemed appropriate. 

In the Monash code, defining the border between convective and radiative regions is done via a method of \textit{relaxation}, introduced in \citet{Lattanzio86}. This typically results in more efficient TDU than relying on the traditional Schwarzschild boundary definition \citep{Frost96}.  This option is not available in MESA, so instead we use the predictive mixing algorithm \citep{Paxton18instrument4, Constantino15tre, Bossini15unc}. We note that there are two algorithms available in MESA that aid core growth, the second of which is convective premixing. As discussed in \citet{Ostrowski21}, there are benefits and drawbacks to both options. Predictive mixing can result in core splitting, particularly during core helium burning. The radiative gradient develops a local minimum in this phase, which causes the core to split into two convective regions with a radiative region in between. Convective premixing, on the other hand, can result in breathing pulses. There exists no strong consensus on whether these pulses are physical or artificial, so here we prefer to avoid them. We chose to use the predictive mixing algorithm given that we could avoid core splitting by managing the temporal and mesh resolution. We followed the guidance of \citet{Ostrowski21} and switched on \verb|predictive_avoid_reversal = `he4'| and \verb|predictive_superad_thresh = 0.005|.

In the default version of the Monash code---the one used to produce the evolutionary sequences for our previous paper, \citet{Karakas22paper1}---we do not require convective overshoot to generate some third dredge up episodes. We discuss in section 5.1 of \citet{Cinquegrana22paper2} that it is likely that third dredge up occurs at lower initial masses than we find in the Monash code. However, MESA requires some form of overshoot to initiate the mixing process at all. Given that there is no third dredge up in the $2 M_{\odot}$ Monash model \citep{Karakas14He}, we only implement the exponential overshoot scheme, or \textit{convective boundary mixing} of \citet{Herwig00} in the $7 M_{\odot}$ MESA model. We choose the same values for the diffusive overshoot parameters as is discussed in \citet{Pignatari16}.

\subsection{Convective stability criteria}

The convective stability criterion used in Monash is the Ledoux criterion: 

\begin{equation}
    \nabla_{\rm rad} < \nabla_{\rm ad} + \frac{\phi}{\delta} \nabla_{\mu}, 
\end{equation}
as stated in \citet{Kippenhahn12}, where $\nabla_{\rm rad}$ and $\nabla_{\rm ad}$ are the radiative and adiabatic temperature gradients, $\phi$ and $\delta$ are density gradients with respect to composition and temperature, respectively, and $\nabla_{\mu}$ is the composition gradient. The Ledoux criterion does not assume chemical homogeneity and so it allows the user to model semi-convection. Semi-convection is available in MESA as well as the Monash code---in the latter case, in the manner described in \citet{Lattanzio86}---but we do not consider the effects of semi-convection in our calculations here. We also note that there are differences in how the two codes treat entropy, which is directly related to the evaluation of thermodynamic quantities. We direct the reader to \citet{Wood1981} for discussion on the treatment of entropy in the Monash code (and \citealt{Paxton10instrument1} for MESA).

Though MESA includes the option to use the Ledoux criterion as well, we encountered convergence difficulties in our models when attempting use it. We chose instead to use the Schwarzschild criterion,
\begin{equation}
    \nabla_{\rm rad} < \nabla_{\rm ad},
\end{equation}
to which the Ledoux criterion reduces under the assumption of uniform composition. As recently demonstrated in \citet{Anders22schwarzschild}, this difference is not a significant concern.

\subsection{Opacities}

For both Monash and MESA, we use the high-temperature OPAL opacities \citep{Iglesias96}. For the low-temperature opacities, we generated tables using the {\AE}SOPUS tool from \citet{Marigo09} that has been integrated within the most recent version of MESA (MESA Instrument Paper VI, \textit{in prep}). These tables are based on the solar mixture of \citet{Lodders03}, where we use scaled-solar initial compositions. Details on the initial conditions used to generate our tables can be found in the appendix of the arXiv version of \citet{Cinquegrana22solarcal}. We note that the OPAL opacities in the Monash code are based on the solar composition of \citet{Lodders03}. These are not currently available for OPAL in MESA, where the solar scaling of \citet{Grevesse98solar} is assumed. For future work (concerned with surface abundances) we advise the reader to generate molecular opacities with the \citet{Grevesse98solar} scaling for consistency, short of the \citet{Lodders03} OPAL opacities being implemented in MESA.

\subsection{Atmospheres}

In both programs, we model the stellar atmospheres using a \textit{gray} $t$--$\tau$ relation. This is an analytical formulation involving direct integration of pressure with respect to opacity, with, for example, $T(\tau)$ given by
\begin{equation}
\def\tfrac#1#2{{\textstyle \frac{#1}{#2}}}
    T^{4}(\tau) = \tfrac34 T_{\rm eff}^{4} \bigl(\tau + \tfrac23\bigr),
    \label{eq:ttau}
\end{equation}
in the case of the Eddington relation \citep{Kippenhahn12}.

In MESA, this is implemented by setting the \verb|atm_option| to \verb|'T_tau'|. The \verb|atm| module is discussed in detail in \citet{Paxton10instrument1}, and the relevant definitions remain the same in \citet{Paxton19instrument5}.


\subsection{Mass loss}

\begin{figure} 
    \centering
    \includegraphics[width=\columnwidth]{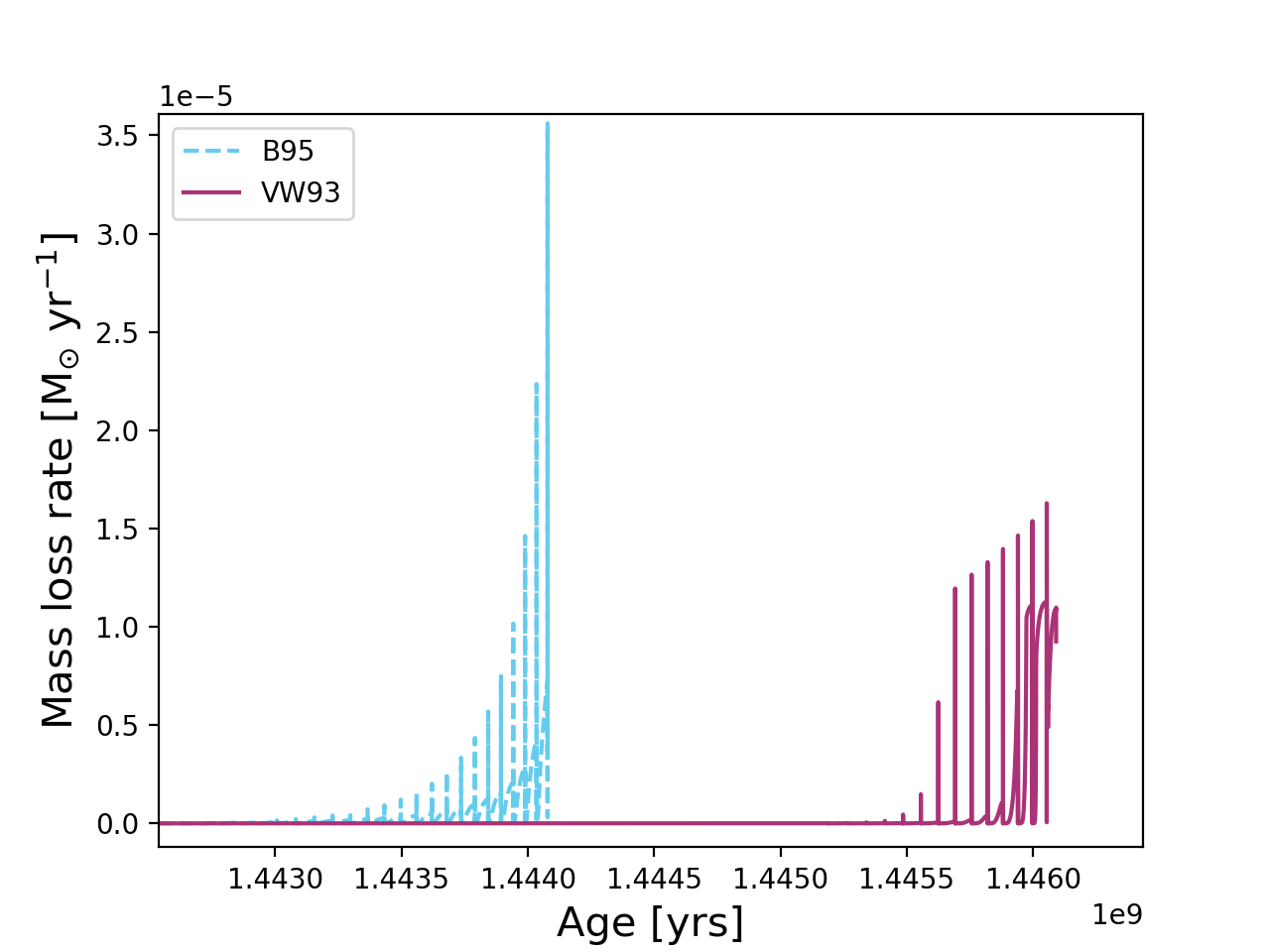}
    \caption{A pair of $2 M_{\odot}$, $Z=0.03$ models both run with the Monash code. The cyan blue (dashed line) model is run using the mass-loss approximation of \citet{Blocker95a}, the magenta (solid line) model with \citet{Vassiliadis93} mass loss.} 
    \label{fig:ML}
\end{figure}

Mass loss on the red giant branch is modelled using the \citet{Reimers75} approximation in both codes, with $\eta _{\rm reimers}=0.477$ assigned in both based on \citet{Mcdonald15mass}. For the AGB, we use the \citet{Blocker95a} approximation given its availability in both codes. We note, however, that in the preceding papers using the Monash code, \citet{Karakas22paper1, Cinquegrana22paper2}, we used \citet{Vassiliadis93} to model mass loss on the AGB. 

We show in Figure \ref{fig:ML} two $2 M_{\odot}$, $Z=0.03$ models, both run with the Monash code but one with \citet{Blocker95a} mass loss, the other \citet{Vassiliadis93}. It is clear that the mass loss rate is faster using \citet{Blocker95a}, which usually results in fewer thermal pulses and therefore a shorter thermally pulsing AGB lifetime than models computed using the prescription of \citet{Vassiliadis93}. The impact of varying the two mass-loss approximations in the Monash code is discussed further in \citet{Karakas16, Karakas18}. 

The scaling factor for the \citet{Blocker95a} approximation, $\eta _{\text{Bl\"{o}cker}}$, is more difficult to match across codes. Given the dependence of \citet{Blocker95a} on radius, mass, and luminosity, small changes in the definition of these variables will produce discrepancies in the rate of mass loss. Setting $\eta _{\text{Bl\"{o}cker}}=0.02$ in MESA produces only two thirds of the thermal pulses that the Monash model endures. Reducing this parameter by half (to $\eta _{\text{Bl\"{o}cker}}=0.01$) in MESA gives the best replication of the core mass growth and envelope loss along the thermally pulsing AGB observed using Monash. To investigate precisely why this variance is required is beyond the scope of this work, but as discussed in \S~\ref{sec:intro}, a high degree of variance in this parameter between codes is typical. 

\section{Results} \label{sec:results}

\begin{figure} 
    \centering
    \includegraphics[width=\columnwidth]{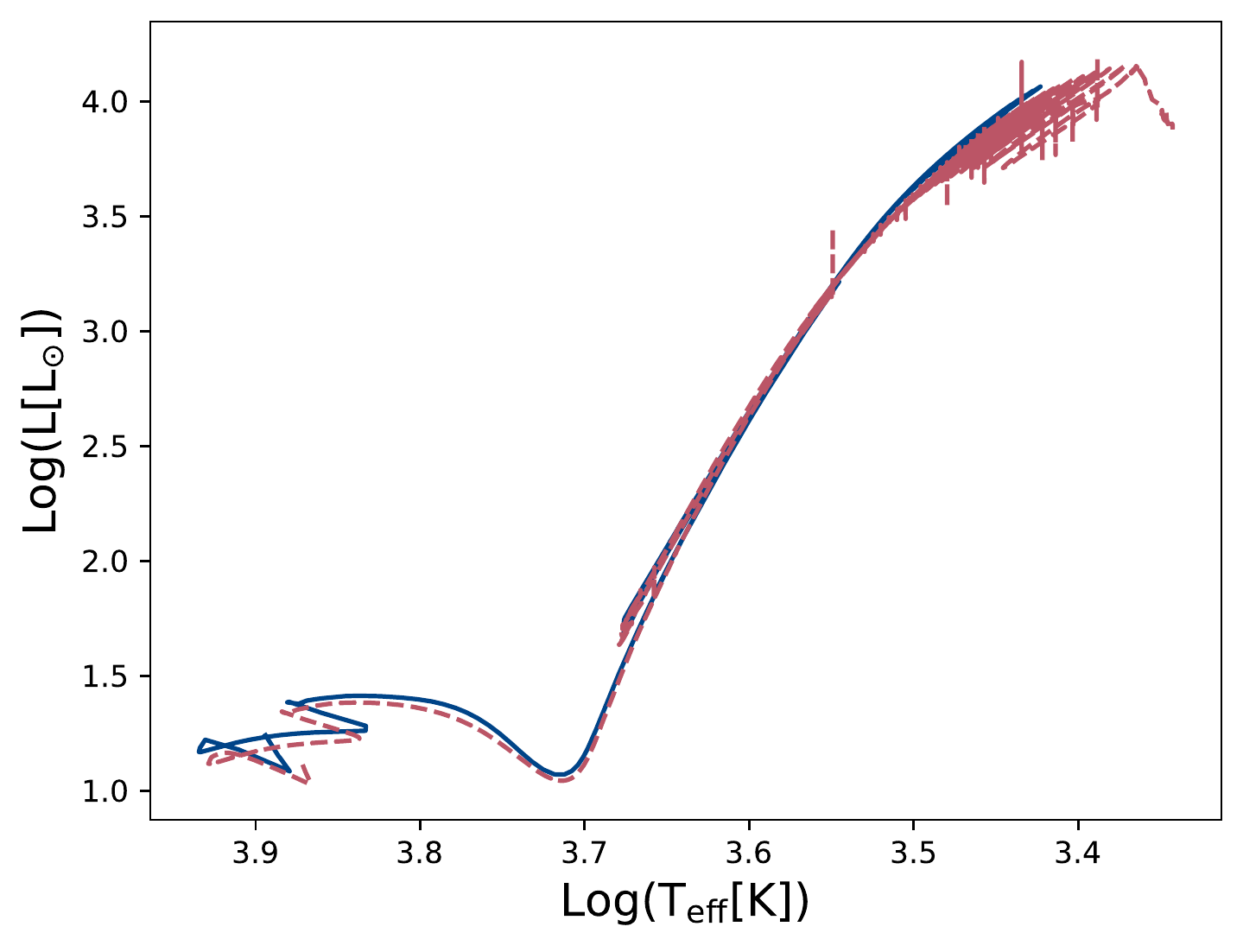}
    \caption{Evolutionary tracks of the two $2 M_{\odot}$, $Z=0.03$ models.} 
    \label{fig:2}
\end{figure}

\begin{figure*} 
    \centering
    \includegraphics[width=\hsize]{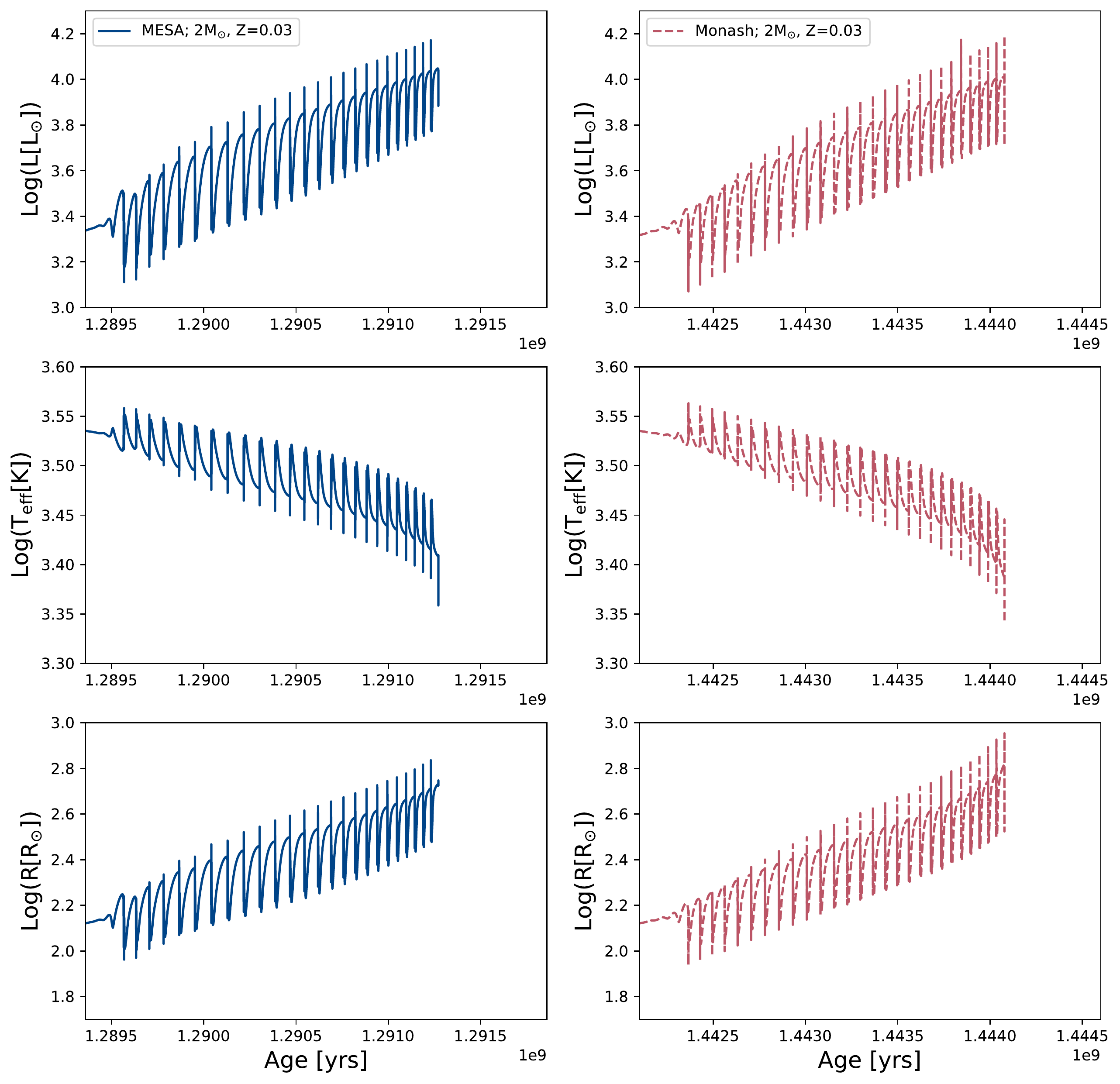}
    \caption{A comparison of the surface properties of the two models. We show luminosity (top panel), effective temperature (center panel) and radius (bottom panel) as functions of time on the thermally pulsing AGB. All show excellent agreement between the models.} 
    \label{fig:3}
\end{figure*}

\begin{figure*} 
    \centering
    \includegraphics[width=\hsize]{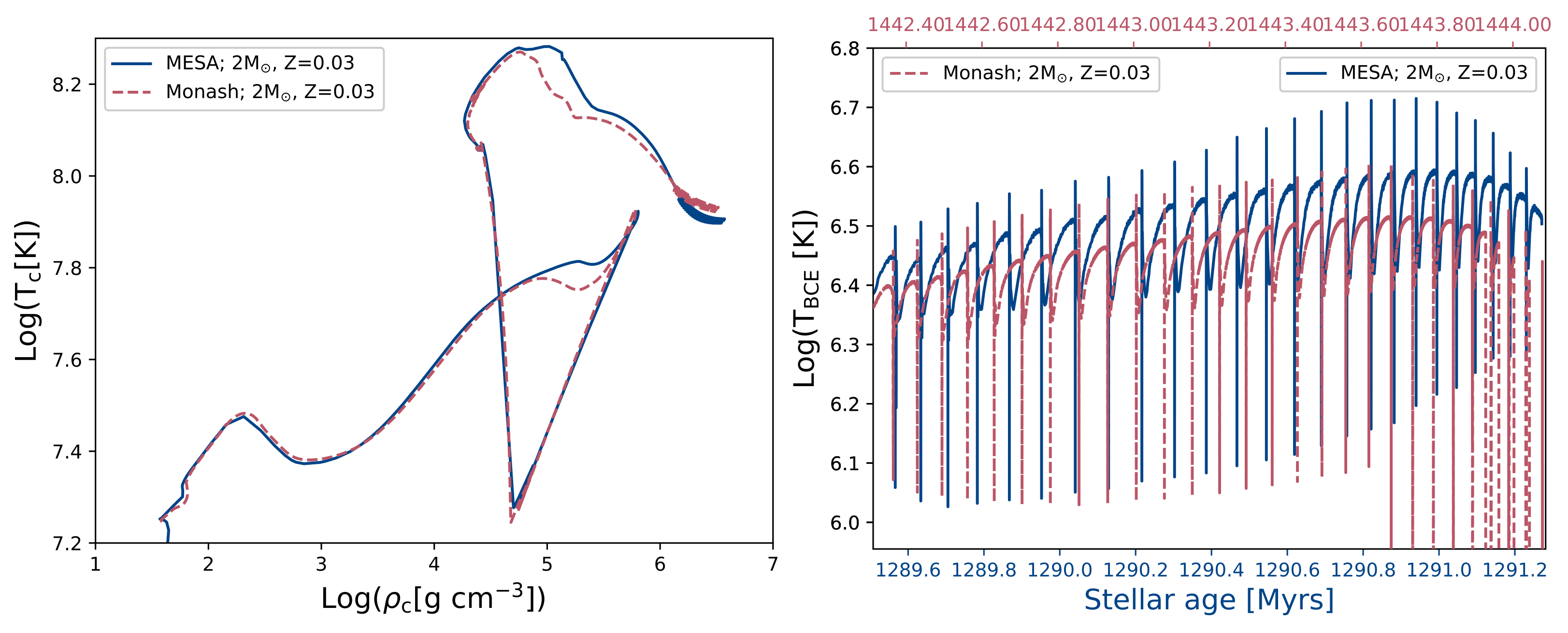}
    \caption{ Left panel: central temperature versus central density. Right panel: maximum temperatures occurring at the base of the convective envelope (T$_{\rm BCE}$), on the thermally pulsing AGB. Variation in elemental abundance is heavily dependent on uncertainties in opacity, mixing and atmospheric boundary conditions, which we have not focused on replicating identically. Here, we show that our central conditions are almost identical, and so any potential variance in surface abundances is due to these mixing uncertainties, rather than differences in the conditions for nuclear burning. We note that the extreme variations in luminosity (appearing as sharp vertical lines) observed in the Monash model during the AGB are numerical artifacts; these are also present in Figure \ref{fig:2}. They are the result of convergence issues, driven by rapid, high-amplitude variations of the luminosity. Such artifacts are typical of most calculations in this regime and are often removed by hand. It is important to note, however, that MESA does not show these variations at the same resolution. In short, this is because MESA was designed with convergence as the primary goal, but this success carries the consequence of much longer run times.}
    \label{fig:4}
\end{figure*}

\begin{figure*} 
    \centering
    \includegraphics[width=\hsize]{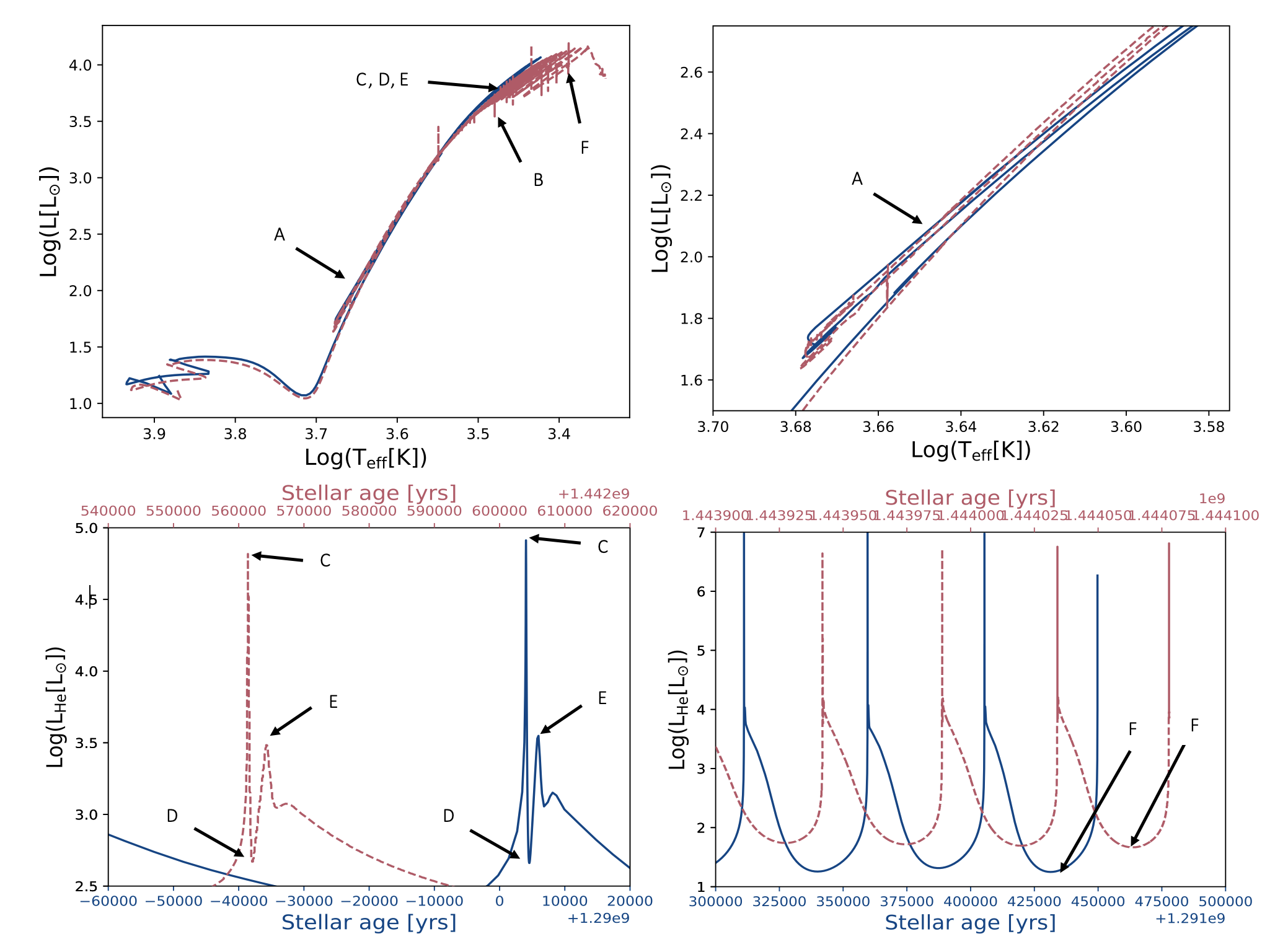}
    \caption{Here we define the evolutionary points at which we compare the models numerically. In the top left panel, we show where points A through F fall in the context of the global evolutionary sequence. Physically, the points are defined as follows:  
    A denotes the termination of core helium burning;
    B is measured at the luminosity maximum of the first thermal pulse;
    C, D, and E are taken at the local luminosity minimum, luminosity maximum, and median of the interpulse quiescent region for the fourth thermal pulse; 
    and 
    F is taken at the local luminosity minimum of the final thermal pulse, regardless of pulse number (i.e. if the MESA model experiences 26 pulses but the Monash model experiences 28, F is measured at pulse 26 for MESA and pulse 28 for Monash.). The top right panel zooms in to show the core helium burning phase, where A is defined. The bottom two panels show where C, D, E, and F fall during the thermally pulsing phase.
    }
    \label{fig:5}
\end{figure*}

\begin{figure*} 
    \centering
    \includegraphics[width=\hsize]{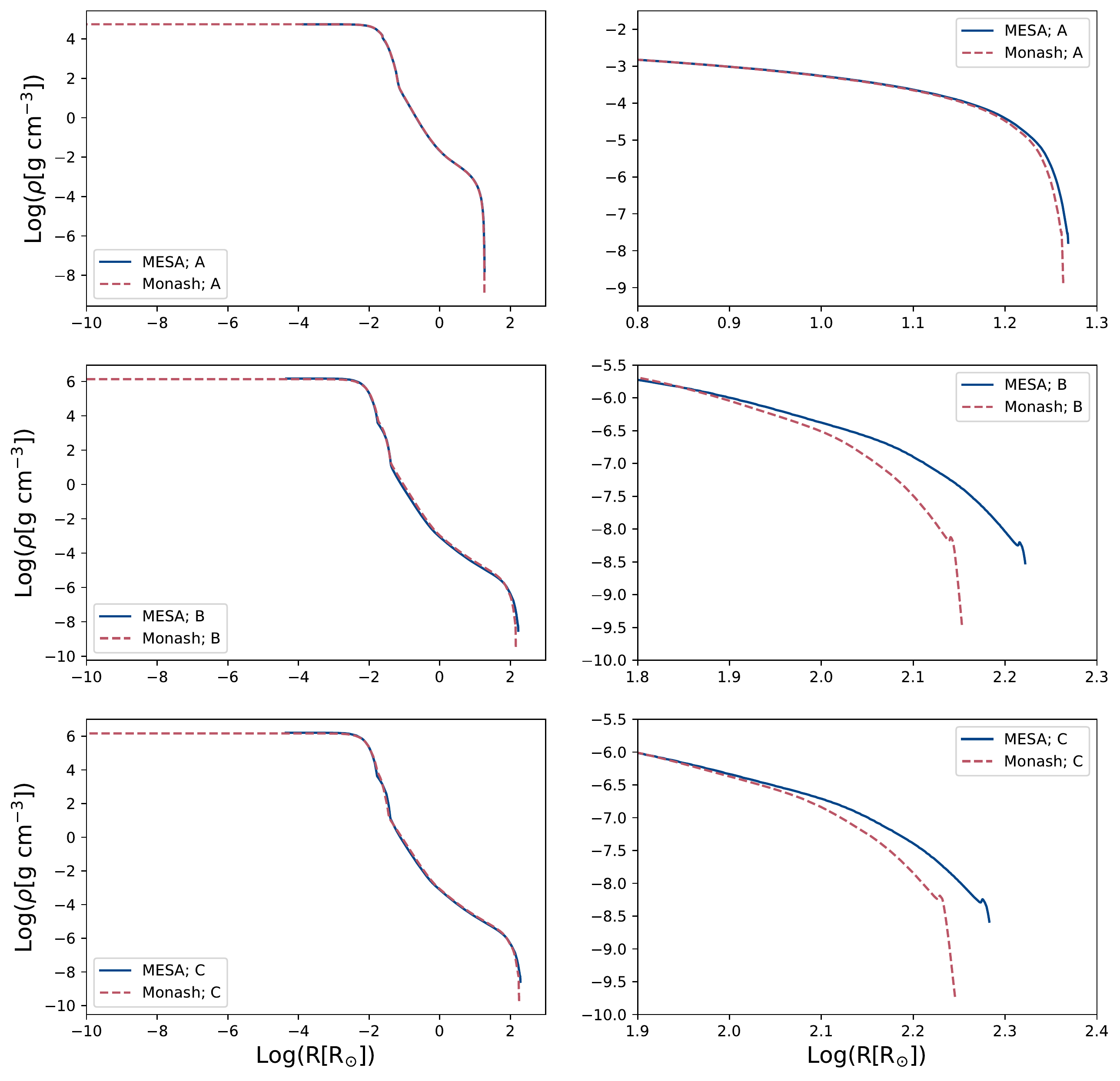}
    \caption{Comparison of the density profiles between the $2M_{\odot}, Z=0.03$ models. The panels on the left show density as a function of radius for given points in evolution: A (top), B (center) and C (bottom). The panels on the right show a closer version of the same plots, identifying the small deviations in density in the outermost region of the envelope.} 
    \label{fig:6a}
\end{figure*}

\begin{figure*} 
    \centering
    \includegraphics[width=\hsize]{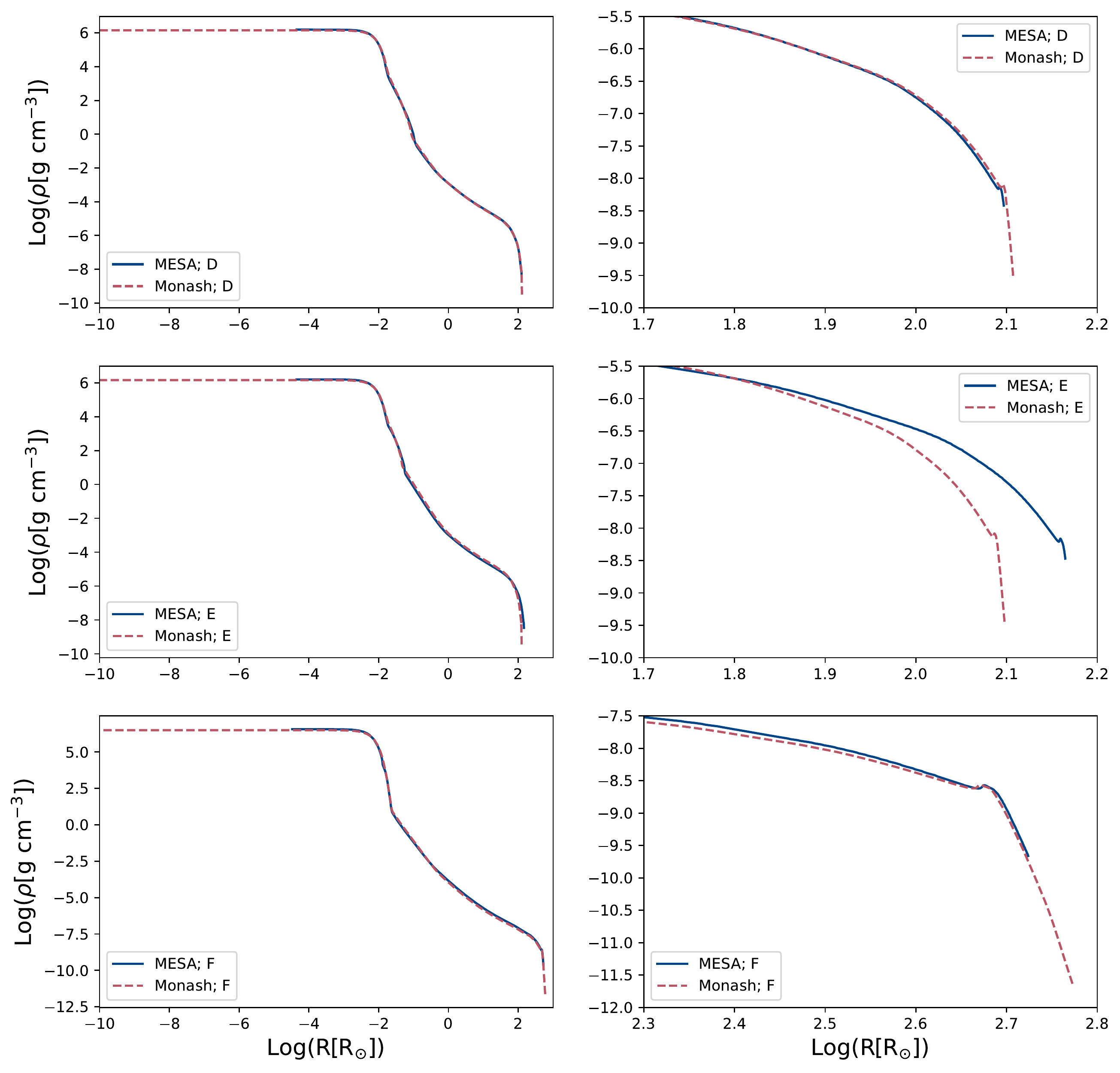}
    \caption{Same as for Figure \ref{fig:6a}, but for points D, E and F.} 
    \label{fig:6b}
\end{figure*}

\begin{table*}
\begin{center}
\caption{Quantitative comparison of variables at each point in evolution. M${}=2$M$_{\odot}$, Z${}=0.03$.}
\label{table:1}
\begin{tabular}{|l|cc|cc|cc|cc|cc|cc|} \hline \hline
Points & \multicolumn{2}{|c|}{Log(L/L$_{\odot}$)} & \multicolumn{2}{|c|}{Log(R/R$_{\odot}$)} & \multicolumn{2}{|c|}{M$_{\rm c}$ [M$_{\odot}$]} &\multicolumn{2}{|c|}{ M$_{\rm env}$ [M$_{\odot}$]} 
& \multicolumn{2}{|c|}{$\rho$ agreement}  \\
 & Monash & MESA  & Monash & MESA  & Monash & MESA & Monash & MESA & at 20\% Log(R) & at 98\% Log(R)\\
 \hline
A & 2.070 & 2.073 & 1.263 & 1.261 & 0.510 & 0.499 & 1.431 & 1.435 & 0.5\% & 0.4\%\\
B & 3.361 & 3.477 & 2.153 & 2.222 & 0.557 & 0.534 & 1.383 & 1.400 & 2.6\% & 0.7\%\\
C & 3.479 & 3.559 & 2.246 & 2.283 & 0.563 & 0.541 & 1.377 & 1.392 & 1.8\% & 0.9\%\\
D & 3.299 & 3.304 & 2.107 & 2.097 & 0.563 & 0.541 & 1.377 & 1.392 & 0.7\% & 1.5\%\\
E & 3.287 & 3.399 & 2.098 & 2.165 & 0.563 & 0.541 & 1.377 & 1.392 & 2.7\% & 0.5\%\\
F & 4.002 & 4.040 & 2.773 & 2.724 & 0.664 & 0.654 & 0.782 & 0.905 & 2.8\% & 9.9\%\\
\hline \hline
\end{tabular}
\noindent 
\end{center}
\end{table*}

\begin{table*}
\caption{Final quantities; M${}=2$M$_{\odot}$, Z${}=0.03$.}
\label{table:2}
\begin{tabular}{|l|c|c|c|c|c|c|c|c|c|} \hline \hline
 Program & \multicolumn{1}{|c|}{$\tau_{\rm stellar}$} & \multicolumn{1}{|c|}{$\tau_{\rm MS}$} & \multicolumn{1}{|c|}{$\tau_{\rm CHeB}$} & \multicolumn{1}{|c|}{$\tau_{\rm TP-AGB}$} & \multicolumn{1}{|c|}{\#TPs} & \multicolumn{1}{|c|}{M$_{\rm c, f}$} & \multicolumn{1}{|c|}{M$_{\rm env, f}$} & \multicolumn{1}{|c|}{C/O$_{\rm f}$} & \multicolumn{1}{|c|}{Run time}\\
 \multicolumn{1}{|c|}{} & \multicolumn{1}{|c|}{[Myrs]} & \multicolumn{1}{|c|}{[Myrs]} & \multicolumn{1}{|c|}{[Myrs]} & \multicolumn{1}{|c|}{[Myrs]} & \multicolumn{1}{|c|}{} & \multicolumn{1}{|c|}{[M$_{\odot}$]} & \multicolumn{1}{|c|}{[M$_{\odot}$]} &  & \multicolumn{1}{|c|}{[Hrs]}\\
\hline
 Monash & 1444 & 1070 & 9.6 & 1.712 & 28 & 0.669 & 0.590 & 0.27 & $\approx$ 1.5\\
 MESA & 1291 & 988 & 8.6 & 1.734 & 26 & 0.654 & 0.904 & 0.25 & $\approx$ 25.3 \\
\hline \hline
\end{tabular}
\medskip\\
\noindent 
\end{table*}


Here, we present the results of the 2M$_{\odot}$, Z${}=0.03$ model comparison. In Figures \ref{fig:2} and \ref{fig:3}, we assess the agreement of various surface quantities between the two models. There is good agreement between the evolutionary tracks in Figure \ref{fig:2}. The MESA model has a slightly more luminous track as it evolves along the main sequence and ascends the red giant branch. In Figure \ref{fig:3}, we plot luminosity, effective temperature, and radius as functions of time along the thermally pulsing AGB and find that these pulsation spectra are almost visually indistinguishable in all coordinates shown. We do note the difference in age on the $x$-axis; however, it is well established that evolutionary \textit{time} (i.e., age) is not a model-agnostic variable, in contrast with evolutionary \textit{phase} \citep{Dotter16eeps}. The more relevant feature is that the \textit{durations} of time displayed are equal, and so we can see that the interpulse periods show strong agreement. 

A code validation study requires reproductions of the structural output as well, so although the surface properties in Figures \ref{fig:2} and \ref{fig:3} show good agreement, they are not a complete portrait of the consistency between the models. Further, surface quantities are heavily dependent on the treatment of convection, mixing, and the choice of atmospheric boundary conditions---which constitute some of the largest sources of uncertainty in stellar modeling. Regarding atmospheric boundary conditions alone, uncertainties are introduced by both (1) the choice of optical depth at which stellar atmospheres are attached, and (2) the atmospheres then calculated/attached at those points. In both Monash and MESA, the atmospheres are modelled using a \textit{gray} $t$--$\tau$ relation and the atmospheric boundary is defined at the same point: where T$_{\rm surface}={}$T$_{\rm eff}$. Because this is a 1D approximation, it does not capture some additional physics, such as 3D and non-local thermodynamic equilibrium (NLTE) effects. While the the assumption of thermodynamic equilibrium is reasonably valid within the stellar interior, it breaks down in the low temperature and density conditions of atmospheres. Monash uses black body integration, but other examples of gray atmospheres include the T--$\tau$ relations of \citet{Eddington26} (as shown in Equation \ref{eq:ttau}) and \citet{Krishna66}. Gray atmospheres are deemed to be inappropriate for very low mass stars ($< 0.9 M_{\odot}$; \citealt{Chabrier00theory, Feiden12reevaluating, Stassun14empirical}) given the intricacies that accompany molecular environments at high densities, but they are decent approximations for masses greater than $0.9 M_{\odot}$. 


These uncertainties fully explain deviations between the MESA and Monash density profiles in the outermost regions of the envelope. Regardless, we do note these deviations are still small: the maximum deviation in density agreement is less than 10\% at 98\% of LogR.

AGBs are significant polluters of the interstellar medium, and so AGB evolutionary sequences \citep[e.g.][]{Karakas22paper1} are often fed into nucleosynthesis codes \citep{Cinquegrana22paper2}. This allows us to investigate the chemical composition of the integrated mass expelled from the star over its lifetime (i.e., stellar yields). These yields are the input used in galactic chemical evolution studies \citep[see][]{Kobayashi20} to model the chemical contributions of stars at a galactic scale. 

In this work, we do not engage in extensive discussion of the surface abundances of our models for a number of reasons. As discussed in \S~\ref{sec:intro} and earlier in this section, there are numerous uncertainties that impact precise values of surface quantities, and these effects are especially heightened for surface abundances. First of all, they are subject to the same uncertainties in atmospheric boundary conditions mentioned in the previous paragraph. 


Second, the elemental abundances on the surface of the model are not only governed by the nuclear conditions of the stellar interior, but also by the mixing processes that dredge those products up into the envelope. The depth and efficiency of mixing varies between the models (that is, the first dredge up for the 2$M_{\odot}$ model, and also the second and third dredge up events for the 7$M_{\odot}$ model we have included in \S~\ref{section:A}). This is predominantly a function of our choices for (1) the convective stability criterion (Schwarzschild in MESA vs Ledoux in Monash) and (2) the convective boundary placement algorithm (predictive mixing in MESA vs.\ relaxation in Monash). Using convective overshoot, we can force agreement between the models on the location of the convective boundary during the thermally pulsing phase to some extent, but there still remains some variation. Third, in neither case are the effects of heavy element diffusion or thermohaline mixing taken into account, both of which affect the distribution of abundances in the outer layers of evolved stars (for further discussion on the effects of diffusion, see \citealt{Bahcall95solar, Henney95effects, Gabriel97influence, Castellani97heavy, Chaboyer01heavy}; for thermohaline mixing, see \citealt{Charbonnel10thermohaline1, Lagarde11thermohaline2, Lagarde12, Lattanzio15, Angelou15, Henkel2017}). For these reasons, surface abundances of individual species are not a useful diagnostic for agreement between our models. Out of interest, we do include the final C/O ratios for both models in Table. \ref{table:2} and \ref{table:A2}. There is an 8\% difference in the final C/O ratios of the 2$M_{\odot}$ models. These models do not undergo the third dredge up, so C/O is only depleted during the first dredge up.

We do compare the \textit{conditions} for nucleosynthesis, the sole physical dependencies of which are the central density and temperature conditions. For intermediate-mass stars on the AGB, we also care about the temperature profile at the base of the convective envelope, especially in regards to hot bottom burning \citep{Lattanzio1996, Busso99, Herwig05, Nomoto13, Karakas14, Karakas22paper1, Cinquegrana22paper2, Ventura22nuc}. We show in the left panel of Figure \ref{fig:4} ($\log {\rm T}_{\rm c}$ vs.\ $\log\rho _{\rm c}$) that the central temperature and densities are nearly identical between the models. The right panel of Figure \ref{fig:4} demonstrates that the conditions at the base of the convective envelope are also very similar. While the 2$M_{\odot}$ model shown in this Figure does not reach temperatures sufficient for hot bottom burning, the 7$M_{\odot}$ in \S~\ref{section:A} does exceed the minimum requirement of 5$\times 10^7$K for the onset of this process (see Figure \ref{fig:A3}), so it is an important consideration in general. 

Ensuring that the central temperature and density conditions are consistent is also important given that nuclear reaction activation thresholds govern the stellar luminosity. A more productive core hydrogen burning sequence (resulting in higher luminosities on the main sequence) will produce a more massive hydrogen exhausted core. A larger hydrogen exhausted core mass, in turn, results in a lower initial mass threshold for further burning stages and core collapse. Higher luminosities on the AGB will result in higher mass loss rates (especially when using the approximation of \citealt{Blocker95a}), which will reduce the number of thermal pulses and thus the AGB lifetime, likewise reducing the number of opportunities for third dredge up and hot bottom burning episodes. 

We have defined six points in evolution at which we compare the density profiles between the models, labelled in Figure \ref{fig:5}. The first point, A, indicates the time step immediately following the terminal age of core helium burning, which we define as the point where the central helium mass fraction drops below $10^4$. Our second point, B, is defined as the onset of the first thermal pulse. Points C, D and E are taken during the $4^{\text{th}}$ thermal pulse for both models, which tests the consistency of the models at two local extrema during, and once after, the computationally demanding event. Finally, point F is defined at the minimum of the last thermal pulse each model endures. Given that the exact number of thermal pulses undergone by each model differs, we note this point is not defined at the same thermal pulse number (the $25^{\text{th}}$ thermal pulse for MESA vs the $28^{\text{th}}$ for Monash). The density profiles associated to these evolutionary points are shown in Figures \ref{fig:6a} and \ref{fig:6b}. The left panels show the full radial profile (with the same scaling on the $x$-axes); the right panels show a closer version of the outer radial boundary (note, these $x$-axes vary). In the right-side panels, we note the presence of a density inversion in the outer-most layers of the model in both the MESA and Monash profiles. This is a physical phenomenon known to occur in AGB envelopes, caused by a local maximum in the density distribution at the outer edge of the convection--ionization zone (see \citealt{WoodFaulkner1973, Wood1974, BeckerIben1980} for a more detailed discussion of this physics). This feature is also present in AGB density profiles computed with an earlier version of MESA, as shown, for instance, in Figure 2 of \citet{Joyce19density}. 

At point A, the two profiles are indistinguishable before the models ascend the AGB. With the onset of thermal pulses and the further expansion of the envelope, there is some variance in density in the very outermost region of the envelope, with MESA producing a slightly puffier model. This variance is to be expected given that it occurs in the region most impacted by the dominant sources of uncertainty in our models: opacities and atmospheric boundary conditions. We likewise expect these discrepancies to compound over time as the models move through their evolution, which we observe. Yet, regardless of this, they still show excellent agreement. 

In Table \ref{table:1}, we quantify the differences in luminosity, radius, hydrogen exhausted core mass, envelope mass and density profiles between the codes at each evolutionary point. At the end of core helium burning (A), there is less than 0.2\% difference between the luminosity and radius produced by Monash versus MESA. During the thermally pulsing phase, this grows to a 3.4\% difference in the luminosities, and 3.2\% between the radii. The mass of the hydrogen exhausted core shows less then 4.2\% difference over the entirety of the evolution: this difference ranges from 2.2\% at point A, to a maximum variance of 4.2\% at the first thermal pulse, where the Monash model is slightly larger in all cases. The final core masses (at point F) varies only by 1.5\%.

In Table \ref{table:2}, we compare some global features quantitatively for the two models. There is some variation in the duration of each phase. The Monash model has an $\sim$11\% longer overall lifetime, spending 8\% more time on the main sequence and producing only a $\sim$1\% longer thermally pulsing AGB phase. Further variation in the total lifetimes can be attributed to the extra Myr that the Monash model spends in the core helium burning phase. The duration of core helium burning is heavily impacted by the treatment of convection, semi-convection and overshoot (see \citealt{Constantino15tre, Constantino16tre, Constantino17tre} for a detailed study of this phase). The production of more thermal pulses, in Monash's case, can once again be attributed to differences in choices for outer boundaries and their associated uncertainties, as discussed previously. As the Monash model experiences two more thermal pulses than the MESA model, it loses more of its final envelope mass. However, we are most interested in the hydrogen exhausted core mass, which at the termination of the models' evolution shows less than a 1.5\% discrepancy.

\section{Discussion and conclusions}
\label{section:discussionandconclusion}

In this work, we have validated a $2M_{\odot}$, super-solar-metallicity MESA model using the best possible reproduction of the Monash stellar evolution code's physics. Our aim was to produce in MESA a model whose structural and evolutionary output match the results of the Monash code as closely as possible, so that MESA may be used in the future to extend the Monash code's excellent performance on the TP-AGB to higher mass regimes. We focused on ensuring that this model meets the standard set by the Monash code along the thermally pulsing AGB---an evolutionary regime historically difficult to model and subject to large variation across stellar evolution codes.
We have shown that our model satisfies this objective by comparing both the evolutionary tracks and the structural conditions at six key points in evolution. In our comparison of the density profiles, we find excellent agreement between the models for all points in evolution, observing deviation only in the very outer regions of the envelope. We attribute this still relatively minor discrepancy to differences in opacity and atmospheric boundary conditions and their associated uncertainties. Prior to the AGB, there is a less than 0.2\% variation in luminosity and radius between the models, which extends to a maximum of 3.4\% during the thermal pulses. The hydrogen exhausted core mass---a critical indicator that separates intermediate from massive stars---shows a difference of less than 4.2\% throughout the whole evolution, with the final value less than 1.5\%. We do not compare elemental abundances on the surfaces of the models further than the final C/O ratio, given these are highly sensitive to convection and mixing particularities and uncertainties. The final C/O ratios for the $2M_{\odot}$ models differ by less than 8\%, this is indicative of carbon depletion during the first dredge up. They are not massive enough to undergo the second or third dredge ups at this metallicity. We also compare the interior conditions necessary for these nuclear reactions to occur. The central temperatures and densities are almost identical between the two models. Importantly for AGBs, we also look at the temperature at the base of the convective envelope, which governs the onset of hot bottom burning, and likewise find excellent agreement. This further supports our contention that variations in the stellar surface abundances are a function of (potentially subtle) differences in prescriptions for mixing and convection. 

The Monash code has long been considered state-of-the-art in the TP-AGB regime, having been iterated and refined specifically for the purpose of modeling this phase over four decades. It is the veteran code in the literature for calculating thermal pulses, which are among the most computationally demanding and difficult phases of evolution to model for low- and intermediate-mass stars. As such, best practices for modeling this regime have historically been tied to a closed-source code, available to only a handful of researchers. MESA, though a much younger instrument, has become the most widely used stellar structure and evolution program in astrophysics due to its wide modeling capability and open source availability. By providing the first validation of a thermally pulsing AGB model computed with MESA against Monash, we hope to improve the accessibility of detailed thermally pulsing AGB calculations and lay the groundwork for extending Monash's best practices to higher mass regimes. We have made our inlists publicly available. They can be found at: \dataset[10.5281/zenodo.6955314]{\doi{10.5281/zenodo.6955314}}.



\section*{Acknowledgements}

This research was supported by the Australian Research Council Centre of Excellence for All Sky Astrophysics in 3 Dimensions (ASTRO 3D), through project number CE170100013. G.C.C acknowledges financial support through a travel grant provided by the Astronomical Society of Australia. M.J. acknowledges the Lasker Data Science Fellowship, awarded by the Space Telescope Science Institute, which funded a short research residence for G.C.C during which the majority of this manuscript was written. The authors further acknowledge the MESA developers team for fruitful discussion, in particular Warrick Ball, Evan Bauer, Aaron Dotter, Adam Jermyn, and Pablo Marchant. M.J. further acknowledges L\'aszl\'o Moln\'ar for helpful discussion and John Bourke for typesetting and editorial work on this manuscript. We also thank the anonymous referee for their detailed suggestions which helped us to improve our paper.


\bibliography{Giulias_bib}{}
\bibliographystyle{aasjournal}

\appendix

\counterwithin{figure}{section}
\counterwithin{table}{section}

\section{Validation of an intermediate mass AGB case}
\label{section:A}

In this section, we provide a complete replication of the validation performed in \S~\ref{sec:results} above, but for an intermediate mass AGB model with initial mass and metallicity of $7 M_{\odot}$ and $Z=0.03$, respectively. Our results in this section reinforce the strong agreement found between the two programs. 

\begin{figure} 
    \centering
    \includegraphics[width=9cm]{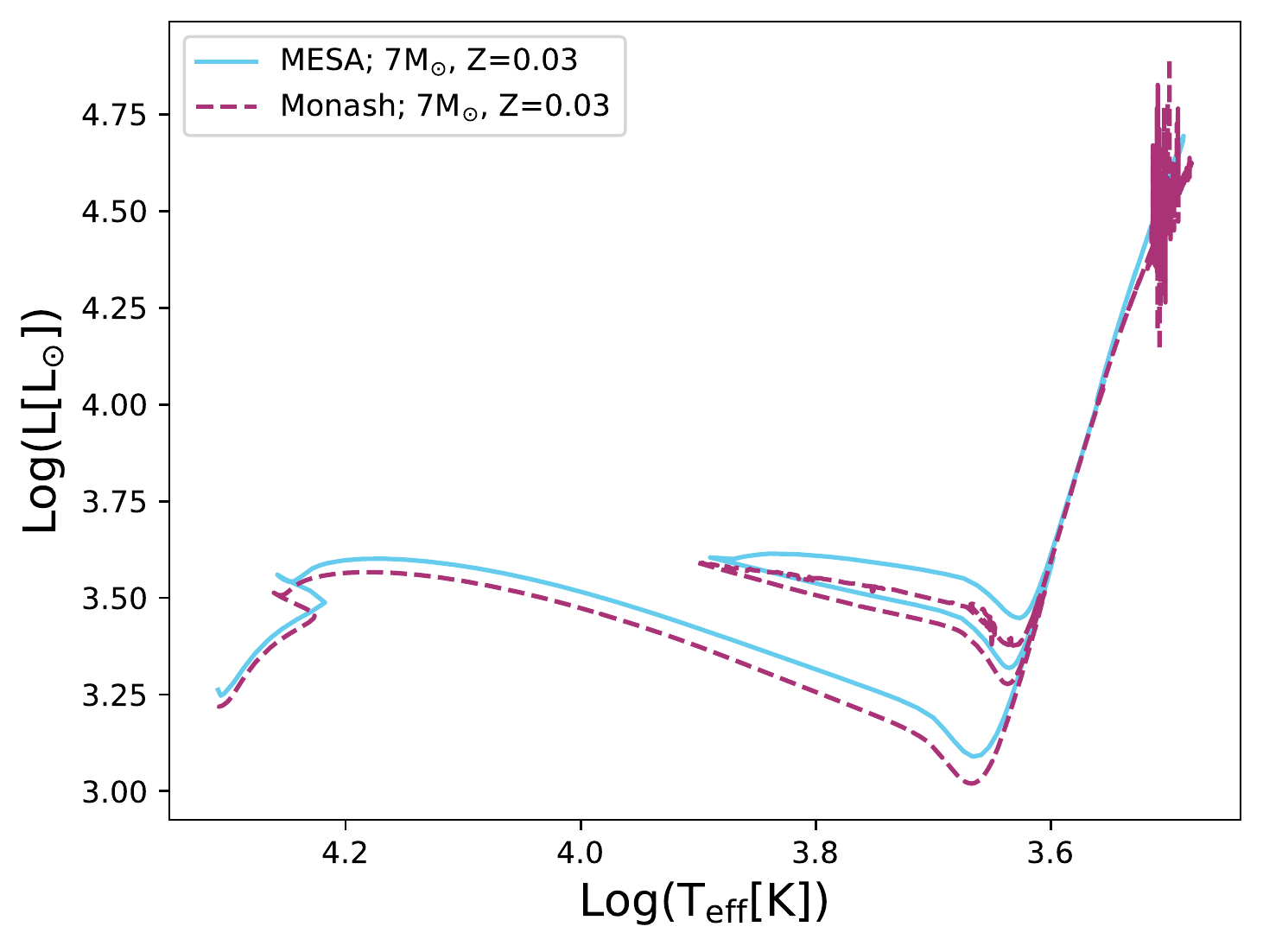}
    \caption{Stellar tracks of the two $7 M_{\odot}$ models.} 
    \label{fig:A1}
\end{figure}

\begin{figure} 
    \centering
    \includegraphics[width=18cm]{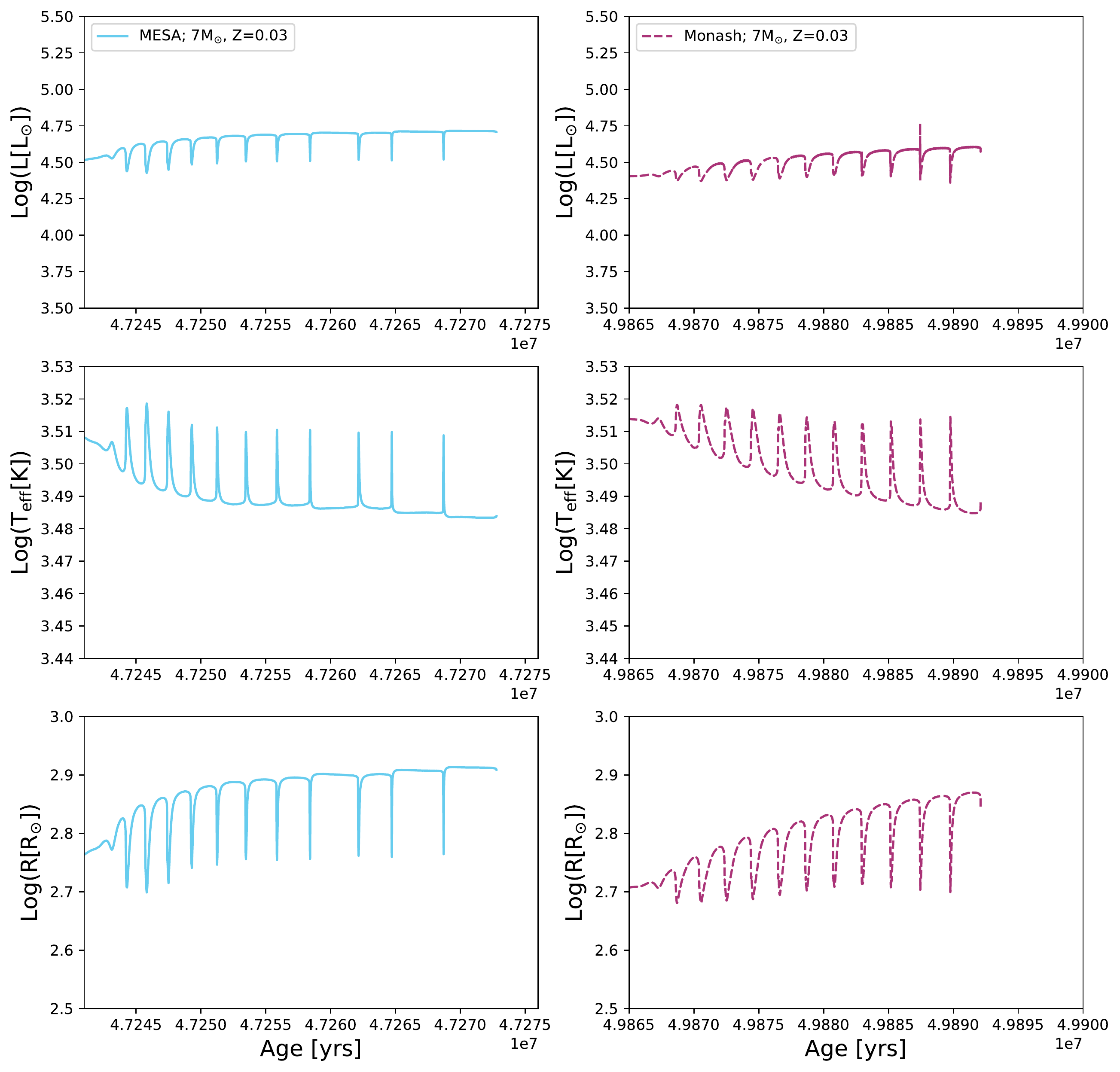}
    \caption{A comparison of the surface properties of the two $7 M_{\odot}$, $Z=0.03$ models. We show luminosity (top panel), effective temperature (center panel) and radius (bottom panel) as functions of time on the thermally pulsing AGB. We note that both evolution runs were halted at the $12^{\text{th}}$ thermal pulse, which is sufficient to demonstrate our point without excess computing time.} 
    \label{fig:A2}
    \end{figure}

\begin{figure} 
    \centering
    \includegraphics[width=18cm]{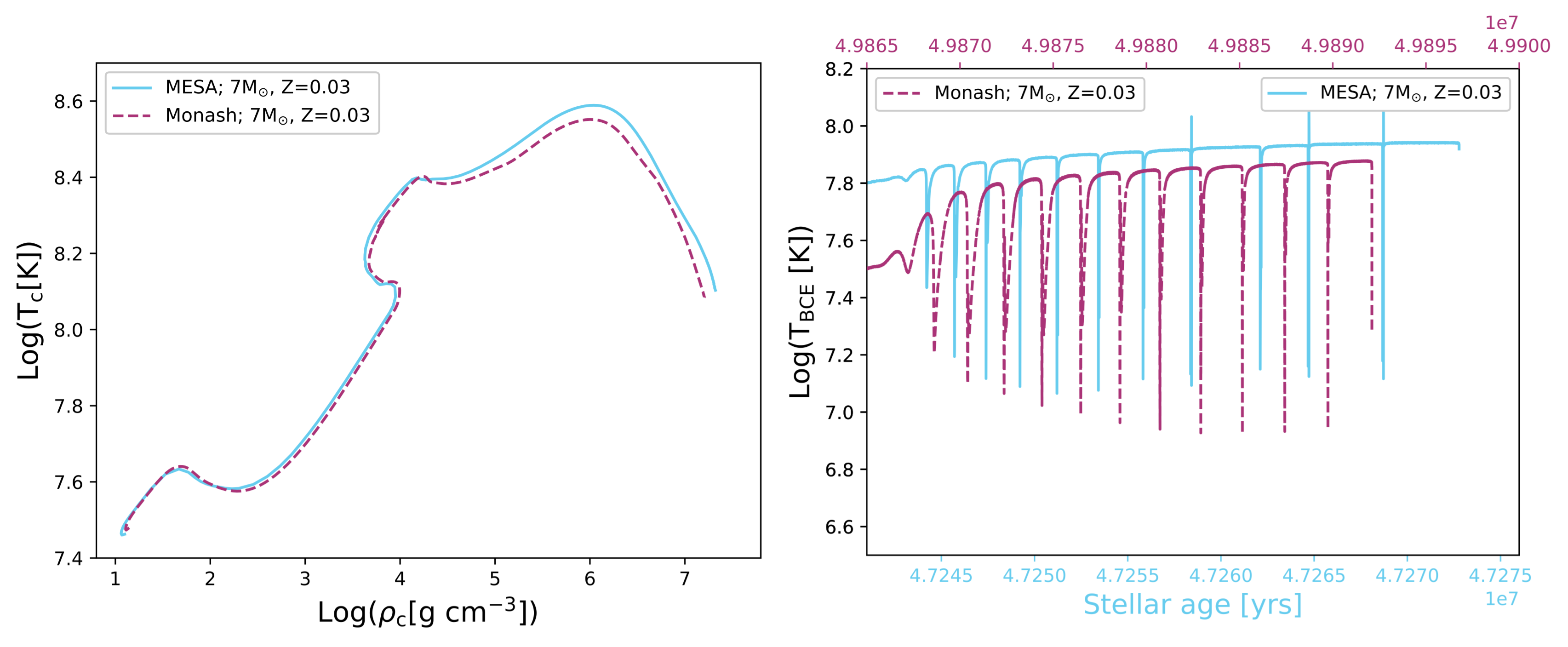}
    \caption{Left panel: central temperature versus central density. Right panel: maximum temperatures occurring at the base of the convective envelope (T$_{\rm BCE}$), on the thermally pulsing AGB. Variance in elemental abundance is heavily dependent on uncertainties in opacity, mixing and atmospheric boundary conditions, which we have not focused on replicating identically. Here, we show that our central conditions are almost identical, and so any potential variance in surface abundances is due to these mixing uncertainties, rather than differences in the conditions for nuclear burning.} 
    \label{fig:A3}
\end{figure}

\begin{figure*} 
    \centering
    \includegraphics[width=\hsize]{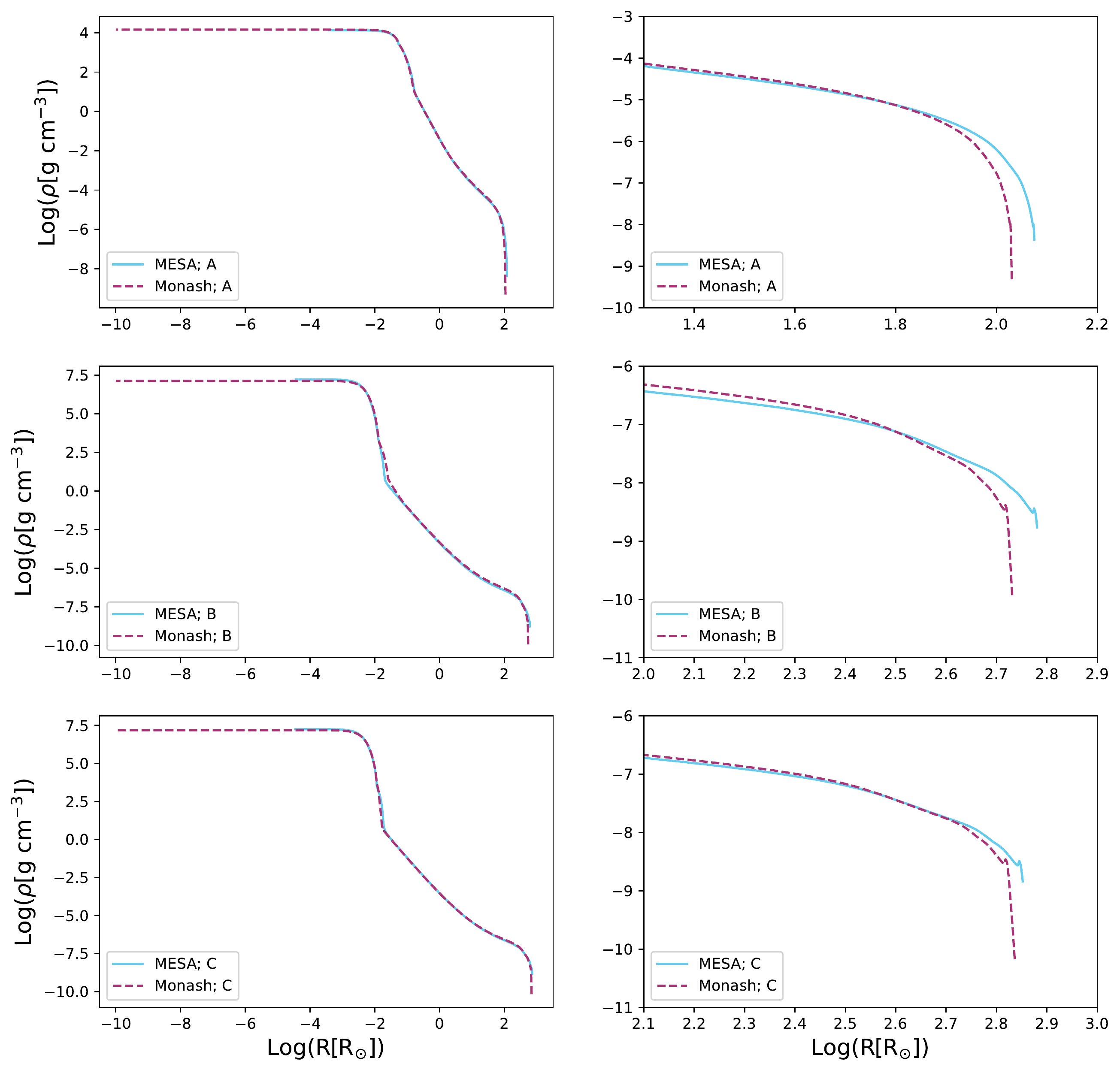}
    \caption{Comparison of the density profiles between the $7M_{\odot}$,  $Z=0.03$ models. The panels on the left show density as a function of radius for given points in evolution: A (top), B (center) and C (bottom). The panels on the right show a closer version of the same plots, identifying the small deviations in density in the outermost region of the envelope.} 
    \label{fig:A5a}
\end{figure*}

\begin{figure*} 
    \centering
    \includegraphics[width=\hsize]{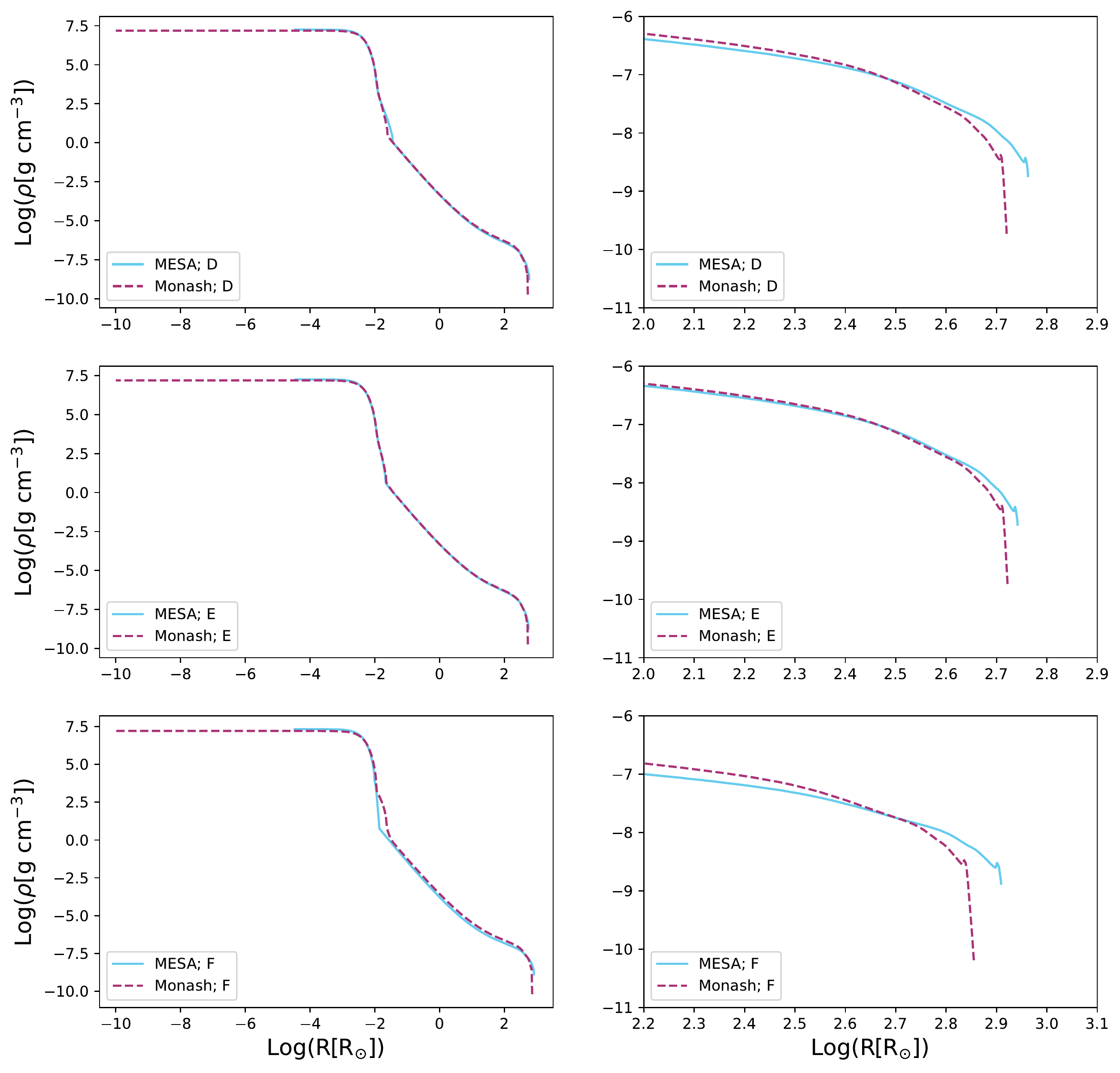}
    \caption{Same as for Figure \ref{fig:A5a}, but for points D, E and F.} 
    \label{fig:A5b}
\end{figure*}

Figures \ref{fig:A1} and \ref{fig:A2}, analogues of Figures \ref{fig:2} and \ref{fig:3} from \S~\ref{sec:results}, show the stellar tracks and various surface quantities (luminosity, effective temperature and radius) as functions of time. The conditions for nuclear burning are exhibited in  Figure \ref{fig:A3} (analogue of Figure \ref{fig:4}). The left panel shows the central temperature as a function of central density; the right panel shows the conditions at the base of the convective envelope during the thermally pulsing AGB. In contrast to Figure \ref{fig:4}, these models reach temperatures sufficient for the onset of hot bottom burning. The density profiles for evolutionary points A to F are contained in Figures \ref{fig:A5a} and \ref{fig:A5b}. Finally, Tables \ref{table:A1} and \ref{table:A2} provide the quantitative data for these points in evolution. We note one significant difference between the $7 M_{\odot}$ and $2 M_{\odot}$ models is found in the final C/O ratio for the $7 M_{\odot}$ case. Here, the Monash model has a 35\% greater final C/O ratio. This model undergoes much more efficient third dredge up, which we did not focus on replicating. To do so, one would need to increase the amount of convective overshoot employed during the thermally pulsing AGB.

\begin{table*}
\begin{center}
\caption{Quantitative comparison of variables at each point in evolution. $M=7M_{\odot}$, $Z=0.03$.}
\label{table:A1}
\begin{tabular}{|l|cc|cc|cc|cc|cc|cc|} \hline \hline
Points & \multicolumn{2}{|c|}{Log(L/L$_{\odot}$)} & \multicolumn{2}{|c|}{Log(R/R$_{\odot}$)} & \multicolumn{2}{|c|}{M$_{\rm c}$ [M$_{\odot}$]} &\multicolumn{2}{|c|}{M$_{\rm env}$ [M$_{\odot}$]} 
& \multicolumn{2}{|c|}{$\rho$ agreement}  \\
 & Monash & MESA  & Monash & MESA  & Monash & MESA  & Monash & MESA & at 20\% Log(R) & at 98\% Log(R)\\
 \hline
A & 3.464 & 3.538 & 2.030 & 2.075 & 1.450 & 1.495 & 5.495 & 5.445 & 1.6\% & 1.9\% \\
B & 4.435 & 4.537 & 2.731 & 2.780 & 0.945 & 0.956 & 5.975 & 5.961 & 2.0\% & 0.5\% \\
C & 4.565 & 4.631 & 2.836 & 2.852 & 0.950 & 0.958 & 5.923 & 5.939 & 1.4\% & 0.3\% \\
D & 4.421 & 4.514 & 2.720 & 2.762 & 0.950 & 0.958 & 5.924 & 5.939 & 1.4\% & 0.5\% \\
E & 4.423 & 4.486 & 2.722 & 2.742 & 0.950 & 0.957 & 5.924 & 5.939 & 0.1\% & 0.5\% \\
F & 4.586 & 4.708 & 2.855 & 2.090 & 0.952 & 0.959 & 5.878 & 5.715 & 5\% & 0.5\%\\
\hline \hline
\end{tabular}
\noindent 
\end{center}
\end{table*}

\begin{table*}
\begin{center}
\caption{Final quantities at 12th TP; $M=7M_{\odot}$, $Z=0.03$.}
\label{table:A2}
\begin{tabular}{|l|c|c|c|c|c|c|c|c|c|} \hline \hline
 Program & \multicolumn{1}{|c|}{$\tau_{\rm stellar [12thTP]}$} & \multicolumn{1}{|c|}{$\tau_{\rm MS}$} & \multicolumn{1}{|c|}{$\tau_{\rm CHeB}$} & \multicolumn{1}{|c|}{$\tau_{\rm TP-AGB [12thTP]}$} & \multicolumn{1}{|c|}{\#TPs} & \multicolumn{1}{|c|}{M$_{\rm c, f}$} & \multicolumn{1}{|c|}{M$_{\rm env, f}$} & \multicolumn{1}{|c|}{C/O$_{\rm f}$} & \multicolumn{1}{|c|}{Run time}  \\
 \multicolumn{1}{|c|}{} & \multicolumn{1}{|c|}{[Myrs]} & \multicolumn{1}{|c|}{[Myrs]} & \multicolumn{1}{|c|}{[Myrs]} & \multicolumn{1}{|c|}{[Myrs]} & \multicolumn{1}{|c|}{} & \multicolumn{1}{|c|}{[M$_{\odot}$]} & \multicolumn{1}{|c|}{[M$_{\odot}$]} & & \multicolumn{1}{|c|}{[Hrs]} \\
\hline
 Monash & 49.9 & 37.8 & 9.6 & 0.0217 & 12 & 0.952 & 5.878 & 0.18 & $\approx$ 2 \\
 MESA & 47.3 & 37.4 & 8.6 & 0.0288 & 12 & 0.959 & 5.72 & 0.035 & 46 \\
\hline \hline
\end{tabular}
\medskip\\
\noindent 
\end{center}
\end{table*}

\end{document}